%% file: dellweg.tex
\documentclass[prl,showpacs,twocolumn,floatfix]{revtex4-1}

\usepackage[utf8]{inputenc}
\usepackage{graphicx}
\usepackage{amsmath, amsfonts, amssymb, bm}
\usepackage{xcolor}

\usepackage{todonotes}
\usepackage{soul}

\renewcommand{\section}[1]{\textit{#1}.---}
\newcommand{\mat}[1]{\mathbf{#1}}

\newcommand{\ket}[1]{| {#1} \rangle}

\date{\today}

\begin{document}
 \title{Spin-polarizing interferometric beam splitter for free electrons}
 \author{Matthias M. Dellweg}
 \author{Carsten M\"uller}
 \affiliation{Institut f\"ur Theoretische Physik I, Heinrich Heine Universit\"at D\"usseldorf, Universit\"atsstr. 1, 40225 D\"usseldorf, Germany}

 \begin{abstract}
  A spin-polarizing electron beam splitter is described which relies on an arrangement of linearly polarized laser waves of nonrelativistic intensity.
  An incident electron beam is first coherently scattered off a bichromatic laser field,
  splitting the beam into two portions, with electron spin and momentum being entangled.
  Afterwards, the partial beams are coherently superposed in an interferometric setup formed by standing laser waves.
  As a result, the outgoing electron beam is separated into its spin components along the laser magnetic field,
  which is shown by both analytical and numerical solutions of Pauli's equation.
  The proposed laser field configuration thus exerts the same effect on free electrons as an ordinary Stern-Gerlach magnet does on atoms.
 \end{abstract}

 \pacs{03.75.-b, 41.75.Fr, 42.25.Ja, 42.50.Ct}

 \maketitle

 \section{Introduction}
 \label{sec:introduction}
  Spin-polarized electron beams are useful for a number of applications in various fields of physics \cite{Kessler1985}.
  They are used, for example, to study magnetic properties in condensed matter systems \cite{Getzlaff2010},
  electron-exchange processes in atomic collisions \cite{Gay2009}, and the inner structure of the nucleon in deep-inelastic scattering \cite{[{See e.g., }]Abe1995,*Alexakhin2007}.
  In high-energy physics, spin-polarized electron (and positron) beams can enhance the experimental sensitivity and render additional observables accessible \cite{Moortgat-Pick2008}.

  There are various methods to generate spin-polarized electrons \cite{Kessler1985}.
  At first sight, the most straightforward way would be to split an electron beam into its spin components, just as a Stern-Gerlach setup does with a beam of atoms.
  However, for charged particles, the spin-separating mechanism in an inhomogeneous magnetic field is hindered by the influence of the Lorentz force,
  as already pointed out by Bohr and Pauli
  \cite{Mott1965,[{For further discussion, see }]Batelaan1997,*Rutherford1998,*Rutherford1998a,McGregor2011}.
  Instead, spin-polarized electrons are produced, for instance, by elastic scattering from unpolarized high-$Z$ atomic targets.
  Due to spin-orbit coupling, considerable degrees of polarization are attainable this way -- though at moderate intensities.
  Alternatively, one may exploit the spin-orbit interaction in bound states and photoionize polarized atoms.
  Of great practical importance is photoelectron emission from GaAs photocathodes \cite{Kessler1985} or,
  more advanced, strained semiconductor superlattices \cite{Maruyama1991, *Mamaev2008} after selective photoexcitation into the conduction band.

  In principle, macroscopic laser fields can also affect the electron spin.
  While laser-electron interactions usually are dominated by the coupling of the field to the electron charge,
  under suitable conditions the electron spin may play a role \cite{Walser2002}.
  Spin effects have theoretically been predicted, for instance, in strong-field photoionization of atoms
  \cite{Faisal2004,Klaiber2014,*Klaiber2014a,*Yakaboylu2015,Barth2013,*Barth2014,Milosevic2016} and,
  very recently, also observed in experiment for the first time \cite{Hartung2016}.
  Besides, spin-flip transitions were studied theoretically in laser-assisted Mott \cite{Szymanowski1998,Panek2002}
  and multiphoton Compton scattering \cite{Ivanov2004,Boca2012,Krajewska2014}.
  In general, laser-induced spin effects were found to be rather small, unless the field frequency or intensity is very high.

  In the present paper, we describe a new method to generate spin-polarized electron beams.
  It relies on coherent electron scattering from laser fields and quantum pathway interferences.
  In the ideal case, the setup is capable of perfectly splitting an incident electron beam into its spin components along the laser magnetic field direction.
  Thus, the field configuration acts as a Stern-Gerlach device for free electrons (see Fig.~1).

  \begin{figure}[b]
   \includegraphics[width=\columnwidth]{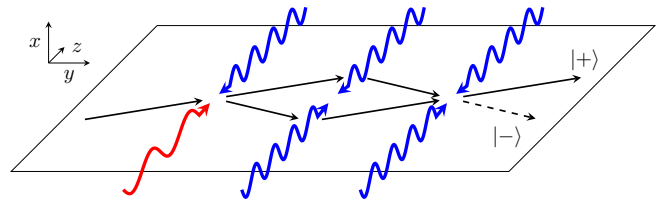}
   \caption{\label{fig:skizze}
    Scheme of the spin-polarizing interferometric beam splitter.
    An incident electron beam is first coherently Bragg scattered off a bichromatic laser field with frequencies $\omega$ (red) and $2\omega$ (blue),
    splitting the beam into two portions.
    Afterwards, the latter are coherently superposed via scattering from monochromatic standing laser waves.
    Due to quantum interference, the outgoing electron beam is separated into its spin components along the laser magnetic field.
    Further details of the beam geometry are specified in the text.
   }
  \end{figure}

  Coherent electron scattering through the Kapitza-Dirac (KD) effect on the periodic potential generated by laser waves resembles the diffraction of light on a grating,
  but with the roles of light and matter interchanged \cite{Kapitza1933,Fedorov1981,Batelaan2007}.
  In its original version \cite{Kapitza1933} the effect involves two photons from a standing wave:
  The electron absorbs a photon of momentum $-\hbar\vec \kappa$
  and emits another of momentum $\hbar\vec \kappa$ (stimulated Compton scattering).
  This way, the electron is elastically scattered, reverting its longitudinal momentum from $+\hbar \kappa$ to $-\hbar \kappa$
  (in our case $\kappa = 2 k$).
  The effect has been confirmed experimentally both in the Bragg \cite{Freimund2002} and diffraction \cite{Freimund2001} regimes.
  Related experiments observed the KD effect on atoms \cite{Gould1986,*Martin1988,*Bucksbaum1988,*Eilzer2014}.

  KD scattering can be sensitive to the electron spin \cite{Ahrens2012,*Ahrens2013,McGregor2015,Dellweg2016}.
  The spin-dependent version,
  used in the first stage of Fig.~1, relies on a three-photon process
  in a bichromatic laser field composed of a fundamental frequency $\omega$ and a counterpropagating second harmonic $2\omega$ \cite{McGregor2015,Dellweg2016}.
  By absorbing one $2\omega$-photon and emitting two $\omega$-photons,
  the energy-momentum balance is fulfilled for Bragg scattering of incident electrons with longitudinal momentum $2\hbar k$.
  The interaction may be considered as arising from an $\vec{A}^2$ term in the Hamiltonian,
  in combination with a $\vec{\sigma} \cdot \vec{B}$ term (which, in general, has to compete with the spin-preserving $\vec{p} \cdot \vec{A}$ term \cite{Smirnova2004}).
  When the incident electron momentum has no component along the field polarization,
  the three-photon process is rendered possible only by the nonzero spin of the electron.
  The latter thus attains a crucial role which is exploited here.
  Note that the spin-dependent KD processes in \cite{Ahrens2012,*Ahrens2013,McGregor2015,Dellweg2016} always yield ``symmetric'' spin effects,
  i.e.\ the spin-flip probabilities up$\to$down and down$\to$up coincide.
  Thus, from an unpolarized incident electron beam, unpolarized outgoing beams result, contrary to the Stern-Gerlach effect.

 \section{Theoretical framework}
 \label{sec:theory}
  Nonrelativistic quantum dynamics of electrons, including their spin degree of freedom, is governed by Pauli's equation.
  In the presence of an electromagnetic field, described by a vector potential $\vec{A}$ in radiation gauge, it reads
  \begin{equation}
   i \hbar \partial_t \psi = \frac{1} {2 m} \left( -i \hbar \vec{\nabla} + \frac{e}{c} \vec{A} \right)^2 \psi + \frac{e \hbar}{2 m c} \vec{\sigma} \cdot \vec{B} \psi
  \label{eqn:pauli}
  \end{equation}
  where $\psi$ is the electron wave function as a Pauli spinor, $m$ the electron mass and $-e$ its charge.
  $\vec{\sigma} = (\sigma_x, \sigma_y, \sigma_z)$ denotes the 3-vector of Pauli matricies.

  The monochromatic KD effect can be formulated by a vector potential for a standing wave in the form
  \begin{equation}
   \vec{A}_\mathrm{m}(t, z) = f(t) a_0 \vec{e}_x \cos \left( 2 \omega t \right) \cos \left( 2 k z + \frac{\chi}{2} \right) \, ,
  \label{eqn:vector_potential_mono}
  \end{equation}
  with amplitude $a_0$, wave number $2 k$ and frequency $2 \omega = 2 c k$.
  The phase parameter $\chi$ allows to adjust the positions of the field nodes.
  A slowly varying envelope function $f(t)$ is introduced to model switching on and off of the laser field.
  In contrast, the bichromatic spin-dependent KD effect relies on electron scattering from two counterpropagating linearly polarized waves,
  \begin{equation}
   \vec{A}_\mathrm{b}(t, z) = f(t) \vec{e}_x \left[ a_1 \cos \left( \omega t - k z \right) + a_2 \cos \left( 2 \omega t + 2 k z \right) \right] \, ,
  \label{eqn:vector_potential_bi}
  \end{equation}
  with frequencies $\omega$ and $2 \omega$.
  Here and henceforth, the incident electron momentum $\vec{p}$ is assumed to lie in the $y$-$z$-plane, being orthogonal to the laser polarization, with $p_z=2 \hbar k$.
  We shall solve Eq.~\eqref{eqn:pauli} for the field configuration of Fig.~\ref{fig:skizze} both by analytical methods and direct numerical integration.

 \section{Analytical treatment}
  Having only $z$-dependence in the potentials, Pauli's equation becomes effectively one-dimensional in space.
  Moreover, by taking a temporal average of the Pauli Hamiltonian for the monochromatic vector potential \eqref{eqn:vector_potential_mono} (with $f \equiv 1$)
  and neglecting constant terms, one obtains a ponderomotive potential \cite{Batelaan2007,Dellweg2015}
  \begin{equation}
   V_\textrm{m} (z) = \frac{e^2 a_0^2}{8 m c^2} \cos \left( 4 k z + \chi \right) \mat{1} \, .
  \label{eqn:mono_pot}
  \end{equation}
  It represents a periodic grating from which the electron beam diffracts.
  Similarly, by means of a Magnus expansion to third order of the Pauli Hamiltonian with the bichromatic vector potential \eqref{eqn:vector_potential_bi},
  \footnotetext{
   For details, see also the Supplemental Material [url], which includes Ref.~\cite{Magnus1954}.
  }\edef\fnsmref{Note\thefootnote}
  one finds a spin-dependent effective ponderomotive potential \cite{Dellweg2016,\fnsmref}\nocite{Magnus1954}
  \begin{equation}
   V_\textrm{b} (z) = - \frac{e^3 a_1^2 a_2 \hbar \omega}{2 m^3 c^6} \sin(4 k z) \sigma_y \, .
  \label{eqn:bi_pot}
  \end{equation}

  The electron wave function
  $\psi(t, z) = \sum_{n \in \mathbb{Z}} c_n(t) e^{i n k z}$
  may be expanded into momentum eigenstates.
  The time evolution is encoded in the Pauli spinors
  $c_n(t) = \left[ \begin{matrix} c_n^\uparrow(t) \\ c_n^\downarrow(t) \end{matrix} \right]$
  which are quantized along the $z$-axis.

  By imposing energy conservation, the Bragg condition allows only the two momentum eigenmodes with $p_z = \pm 2 \hbar k$  to interact with each other.
  We can therefore reduce the ansatz to
  \begin{equation}
   \psi(t, z) = \sum_{n \in \left\{ -2, 2 \right\}} c_n(t) e^{i n k z} = \left( \begin{matrix} c_{-2}(t) \\ c_{2}(t) \end{matrix} \right) \, .
  \label{eqn:ansatz}
  \end{equation}
  A characteristic laser-driven Rabi oscillation dynamics will occur between the two momentum modes.
  In the relevant four-dimensional subspace, the potentials are represented by the block matricies
  \begin{equation}
   V_\mathrm{m} = \frac{\hbar \Omega_\mathrm{m}}{2} \left( \begin{matrix} \mat{0} & e^{-i\chi} \mat{1} \\ e^{i\chi} \mat{1} & \mat{0} \end{matrix} \right) \ , \quad
   V_\mathrm{b} = i \frac{\hbar \Omega_\mathrm{b}}{2} \left( \begin{matrix} \mat{0} & -\sigma_y \\ \sigma_y & \mat{0} \end{matrix} \right)
  \end{equation}
  with the corresponding Rabi frequencies $\Omega_\mathrm{m} = \frac{e^2 a_0^2}{8 \hbar m c^2}$ and $\Omega_\mathrm{b} = \frac{e^3 a_1^2 a_2 \omega}{2 m^3 c^6}$.
  Since all involved momentum eigenstates share the same kinetic energy, we can remove the latter by a gauge transformation into the interaction picture.

  The interaction times with the laser potentials are chosen to yield a quarter (or half) Rabi cycle in each stage of Fig.~\ref{fig:skizze}.
  With $T_\mathrm{m,b} = \frac{\pi}{2 \Omega_\mathrm{m,b}}$, the corresponding time evolution operators thus read
  \begin{equation}
   U_1
    = \exp \left( -\frac{i}{\hbar} T_\mathrm{b} V_\mathrm{b} \right)
    = \frac{1}{\sqrt{2}} \left( \begin{matrix} \mat{1} & -\sigma_y \\ \sigma_y & \mat{1} \end{matrix} \right)
  \end{equation}
  for a $\frac{\pi}{2}$-pulse with the potential $V_\mathrm{b}$,
  \begin{equation}
   U_2
    = \exp \left( -2 \frac{i}{\hbar} T_\mathrm{m} V_\mathrm{m} \right)
    = \left( \begin{matrix} \mat{0} & -i e^{-i\chi} \mat{1} \\ -i e^{i\chi} \mat{1} & \mat{0} \end{matrix} \right)
  \end{equation}
  for a $\pi$-pulse and
  \begin{equation}
   U_3
    = \exp \left( -\frac{i}{\hbar} T_\mathrm{m} V_\mathrm{m} \right)
    = \frac{1}{\sqrt{2}} \left( \begin{matrix} \mat{1} & -i e^{-i\chi} \mat{1} \\ -i e^{i\chi} \mat{1} & \mat{1} \end{matrix} \right)
  \end{equation}
  for a $\frac{\pi}{2}$-pulse with $V_\mathrm{m}$.

  The desired spin filtering effect is achieved for $\chi=\frac{\pi}{2}$.
  The total time evolution operator, as indicated by Fig.~\ref{fig:skizze}, then becomes
  \begin{equation}
   U = U_3 U_2 U_1
   = \frac{1}{2} \left( \begin{matrix} -\mat{1} - \sigma_y && -\mat{1} + \sigma_y \\ \mat{1} - \sigma_y && -\mat{1} - \sigma_y \end{matrix} \right)\ .
  \label{eqn:spin_filter}
  \end{equation}
  When acting on initially $y$-polarized electron states $\ket{\pm} = \frac{1}{\sqrt{2}} (\ket{\uparrow} \pm i \ket{\downarrow})$ with $p_z=2 \hbar k$, we obtain
  \begin{equation}
   U \left( \begin{matrix} 0 \\ \ket{+} \end{matrix} \right)
   = -\left( \begin{matrix} 0 \\ \ket{+} \end{matrix} \right) \quad \mathrm{and} \quad
   U \left( \begin{matrix} 0 \\ \ket{-} \end{matrix} \right)
   = -\left( \begin{matrix} \ket{-} \\ 0 \end{matrix} \right) \, .
  \end{equation}
  Thus, incident electrons with spin-up along the $y$-axis pass the beam splitter with unchanged momentum,
  whereas spin-down electrons are scattered to the mirrored momentum state $p_z=-2 \hbar k$.
  The density matrix of an unpolarized incident electron ensemble is transformed by $U$ into \cite{\fnsmref}
  \begin{equation}
   \rho_\mathrm{unpol} = \frac{1}{4} \left( \begin{matrix} \mat{1} - \sigma_y && \mat{0} \\ \mat{0} && \mat{1} + \sigma_y \end{matrix} \right) \, .
  \label{eqn:rho_unpol}
  \end{equation}
  Thus, the outgoing electrons are spin-filtered into one half with $p_z = 2\hbar k$ and spin-up (along the $y$-axis),
  and the other half with $p_z = -2\hbar k$ and spin-down.

  We point out that our analytical model does not account for field-induced detuning effects which arise at high laser intensities \cite{Dellweg2016}.

 \section{Numerical Results}
 \label{sec:numeric}
  A real-space simulation of an electron wave packet travelling through the electromagnetic field configuration of Fig.~\ref{fig:skizze} corroborates our analytical considerations.
  Initially, the wave packet has central longitudinal momentum $p_z=400~\frac{\mathrm{eV}}{c}$ and spatial width $0.11~\mu\mathrm{m}$;
  its spin is oriented along the positive $z$-axis ($\ket{\uparrow}$).
  Note that the spin expectation value along the $y$-axis thus vanishes.

\begin{figure}[h]
   \includegraphics{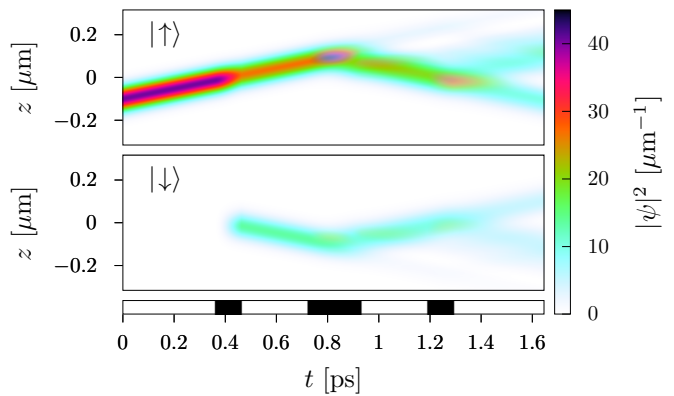}
   \caption{
    Spatial probability density of an electron wave packet with initial central momentum $p_z=400~\frac{\mathrm{eV}}{c}$, width $0.11~\mu\mathrm{m}$ and spin along the positive $z$-axis.
    The projections on $\ket{\uparrow}$ in the upper and $\ket{\downarrow}$ in the lower panel are shown.
    At the bottom, the interaction periods with the lasers are marked.
    The phase parameter $\chi = -\frac{\pi}{10}$ was used.
   \label{fig:wavepacket_z}}
  \end{figure}

\begin{figure}[h]
   \includegraphics{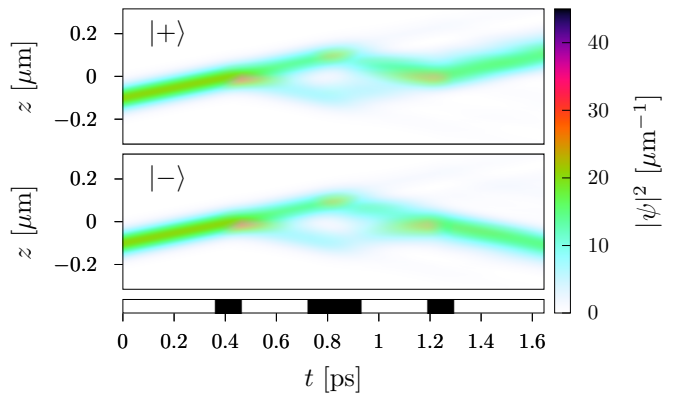}
   \caption{
    Same as in Fig.~\ref{fig:wavepacket_z}, but with the wave function projected on the $\sigma_y$ eigenstates
    $\ket{+} = \frac{1}{\sqrt{2}}\left( \ket{\uparrow} + i \ket{\downarrow} \right)$ in the upper panel,
    and $\ket{-} = \frac{1}{\sqrt{2}}\left( \ket{\uparrow} - i \ket{\downarrow} \right)$ in the lower panel.
   \label{fig:wavepacket_y}}
  \end{figure}

  Figures \ref{fig:wavepacket_z} and \ref{fig:wavepacket_y} show the time evolution of the wave packet, involving different spin projections.
  In the first interaction, a vector potential [see Eq.~\eqref{eqn:vector_potential_bi}] with $e a_1 = e a_2 = 2.35 \times 10^{4}~\mathrm{eV}$
  and $\hbar \omega = 200~\mathrm{eV}$ is switched on
  with $\sin^2$-edges over $5~\mathrm{fs}$%
  for a duration of $106~\mathrm{fs}$.
  Accordingly, the wave packet is partly reflected by the bichromatic KD effect forcing a selective spin flip
  (see lower panel in Fig.~\ref{fig:wavepacket_z}) on the scattered part.
  Thus, after the first interaction, the electron is left in a quantum state where its spin and momentum are entangled
  \footnote{This is sometimes called 'single-particle entanglement', because
  the two entangled q-bits (in the language of quantum information) are encoded on the same particle.}.
  At the second interaction, both partial beams are reflected by a $\pi$-pulse of the type \eqref{eqn:vector_potential_mono} with $e a_0 = 10^{2}~\mathrm{eV}$,
  $\hbar \omega = 200~\mathrm{eV}$ and duration $212~\mathrm{fs}$ without altering their spin state.
  After closing the diamond shape,
  a third KD diffraction (a $\pi/2$-pulse of the same kind as before)
  acts as a spin-insensitive beam splitter to coherently mix the two partial beams in two outgoing momentum channels.

  Figure~\ref{fig:wavepacket_y} illustrates the same time evolution projected on the $\sigma_y$-eigenstates.
  It shows
  that the two output channels of Fig.~\ref{fig:wavepacket_z} are subject to constructive or destructive interference depending on the spin content.
  A strong correlation between momentum and $\sigma_y$-spin of the outgoing states,
  in agreement with Eq.~\eqref{eqn:rho_unpol}, results.
  However, field-induced detuning \cite{Dellweg2016} in the first step leads to uneven splitting of the wavepacket.
  This undesired effect can be compensated for to a large extent by choosing $\chi = -\frac{\pi}{10}$ in the simulation,
  leading to a high polarization degree of the outgoing beams (77~\%).
  Thus, the interplay of various parameters can be exploited to enhance the robustness of the setup against slight imperfections.

 \section{Discussion}
  The results above show that the field configuration of Fig.~\ref{fig:skizze} acts as a Stern-Gerlach magnet for free electrons.
  This is an important finding because it represents a new and methodically uniform solution to the historical problem dating back to Bohr and Pauli
  \cite{Mott1965,[{For further discussion, see }]Batelaan1997,*Rutherford1998,*Rutherford1998a,McGregor2011}.
  A Stern-Gerlach-like device for free electrons purely based on laser waves in an interferometric arrangement has so far not been known.

  In view of an experimental implementation, we note that the single-color KD effect in the Bragg regime,
  as required in the second and third stage of Fig.~\ref{fig:skizze}, has successfully been observed \cite{Freimund2002}.
  The most demanding step is the spin-dependent first stage.
  Nevertheless, despite the rather weak spin interaction, it can be accomplished utilizing nonrelativistic laser intensities. %

  In the example of Figs.~\ref{fig:wavepacket_z} and \ref{fig:wavepacket_y}, the xuv field intensities amount to $I_1 = 7.6 \times 10^{19}~\mathrm{W/cm^2}$ and $I_2 = 4 I_1$,
  corresponding to small values of the relativistic parameter $\xi_{1,2} = e a_{1,2} / mc^2 \approx 0.05$.
  The bichromatic Rabi frequency amounts to $\Omega_\mathrm{b} = \frac{1}{2} \omega \xi_1^2 \xi_2 \approx 10~\mathrm{meV} / \hbar$ ($1.5\times 10^{13}~\mathrm{Hz}$),
  implying a necessary interaction time of $T_\mathrm{b} \approx 0.1~\mathrm{ps}$.
  The latter translates into a laser beam width of $\Delta y = v T_\mathrm{b} \approx 0.3~\mu\mathrm{m}$, where $v$ denotes the transverse component of the electron velocity,
  and also determines the minimum laser pulse duration $\Delta \tau \sim T_\mathrm{b}$.
  An electron energy of 30\,eV is assumed ($v \approx 0.01 c$)
  \footnote{Note that, accordingly, spin-orbit terms which result from the lowest-order relativistic correction to the Pauli Hamiltonian,
  are strongly suppressed by a factor $v/c\approx 10^{-2}$.
  Besides, the correction term $\sim {\vec \sigma}\cdot({\vec A} \times \dot{\vec A})$ vanishes in linearly polarized fields \cite{Erhard2015}.}.
  If the laser beams have spherical cross section with $\Delta x = \Delta y$, we find that the pulse energy is on the order of 50\,mJ.
  Thus, the required laser parameters for the first scattering step are challenging,
  but lie rather close to the performance values of present x-ray free-electron lasers such as the LCLS (Stanford, California).
  Currently, xuv pulses with $\sim 100~\mathrm{fs}$ duration, up to $10^{18}~\mathrm{W/cm^2}$ intensity and few mJ pulse energy are available there \cite{LCLS}.
  Besides, improved x-ray focussing techniques to reach sub-$\mu\mathrm{m}$ beam waists have been proposed \cite{Cornacchia2004,*Schroer2005,*Jarre2005}.
  In light of this, while being demanding, the first stage in Fig.~\ref{fig:skizze} seems to be within experimental reach.
  A supporting conclusion was drawn in \cite{McGregor2015} considering optical laser fields.
  The requirements for the second and third scattering stage are much more relaxed.
  Here the required intensity in the above example is $\sim 10^{16}~\mathrm{W/cm^2}$.
  It could be obtained by outcoupling a small portion from the main laser beam.

  Further requirements exist on the incident electron momentum distribution.
  The optimal longitudinal momentum is $p_z = 2 \hbar k$ to meet the Bragg condition.
  From the width of the Lorentz-shaped resonance curve (see App.~A in \cite{Dellweg2016}) one obtains that the corresponding momentum width is
  $\Delta p_z / p_z = m \Omega_\mathrm{b} / (4 \hbar k^2) =  \xi_1^2 \xi_2 m c / (8 \hbar k)$.
  Thus, for the parameters in our numerical example, $\Delta p_z / p_z \lesssim 0.04$, i.e.\ $\Delta p_z \lesssim 15\,\mathrm{eV}/c$ is required
  (which is fulfilled by $\Delta p_z / p_z \lesssim 0.003$ of the wavepacket in Fig.~\ref{fig:wavepacket_z}).
  The transverse component $p_y$ is responsible for the time $T = m L / p_y$ the electron spends inside the field.
  Here, $L$ denotes the transverse field extent.
  Deviations from the optimum interaction time $T = T_\mathrm{b}$ in the first stage lead to an uncertainty $\Delta P_\mathrm{scatt}$ in the scattering probability
  $P_\mathrm{scatt} = \sin^2 \left( \frac{\Omega_\mathrm{b}}{2} T \right)$ whose ideal value is $\frac{1}{2}$.
  Thus,
  \begin{equation}
   \frac{\Delta P_\mathrm{scatt}}{P_\mathrm{scatt}} = \frac{\pi}{2}\,\frac{\Delta T}{T}
   = \frac{\pi}{2} \left( \frac{\Delta p_y}{p_y} + \frac{\Delta L}{L} \right).
  \end{equation}
  This imperfect beam splitting resulting from $\Delta p_y$ and $\Delta L$ can be partly counteracted by adjusting the phase $\chi$.
  Very crucial is the momentum component along the field direction.
  Deviations $\Delta p_x$ from the ideal value $p_x=0$ will lead to scattering events without spin-flip in the first stage
  due to the ${\vec p}\cdot{\vec A}$ term
  \footnote{For the same reason, a high degree of linear polarization $\gtrsim 99\%$ of the laser beams is necessary,
  which is one order of magnitude better than currently available but can be approached by seeding techniques \cite{Ferrari2015, Allaria2014}.
  Circular-polarized pulses with this level of polarization purity are already available from x-ray free-electron lasers \cite{Lutman2016}.}.
  One can show that the corresponding Rabi frequency is $\Omega_\mathrm{no-flip} = \frac{5 \Delta p_x}{2 \hbar k}\, \Omega_\mathrm{b}$,
  so that $\Delta p_x \ll \hbar k$ is required to suppress these undesired events.
  Finally, if the incident electron beam is spatially broader than the laser beams,
  electrons from the outer beam regions will not interact and, thus, go through undeflected.
  They can be separated from the polarized portions of the outgoing beam by suitable cover plates.

  Laser fields of circular polarization might also appear attractive to coherently control the electron spin
  \cite{Freimund2003,Erhard2015,[{see also }]Bauke2014b,*Bauke2014a}.
  In this case, however, the spin-flip transitions compete with spin-preserving electron scattering because the $\vec{p} \cdot \vec{A}$ interaction term cannot be avoided.
  As a result, the desired spin flips are suppressed.
  Nevertheless, spin filtering of an electron beam, propagating on axis, by coherent scattering from a circularly polarized, monochromatic, standing X-ray wave
  ($\hbar\omega \approx 8$\,keV at $\approx 10^{22}$\,W/cm$^2$) in the weakly relativistic regime has been predicted recently \cite{Ahrens2016}.
  This effect relies on a slight difference between the Rabi frequencies of the two spin orientations along the beam axis.
  By tailoring the interaction time, the scattered beam portion can be extracted spin-polarized.

  Spin-polarization of electron beams may also be achieved by magnetic phase control, exerted either through microscopic current loops \cite{McGregor2011},
  solid-state nano\-structures \cite{Tang2012},
  or solenoids in a Mach-Zehnder interferometer \cite{McGregor2011}.
  The latter proposal combines standing laser waves with static magnetic fields and exploits fine-tuned spin-dependent phase shifts.
  A spin-polarizing Wien filter for electron vortex beams was put forward in \cite{Karimi2012}.
  It relies on multipolar electric and magnetic field geometries with cylindrical symmetry and a conversion of orbital to spin angular momentum
  \cite{[{For recent papers related to the coupling between orbital and spin angular momentum for relativistic electrons, we refer to }] Hayrapetyan2014, *Wen2016}.
  Besides, a spin-insensitive,
  but otherwise similar laser-based interferometric setup as in Fig.~\ref{fig:skizze} is applied to atomic beams by the GAIN project for gravity studies \cite{Schkolnik2015}.
  Also, a spin-insensitive interferometric beam splitter for electrons exploiting the KD effect has been proposed in \cite{Marzlin2013}.

 \section{Conclusion}
 \label{sec:conclusion}
  A new way of generating spin-polarized electron beams has been theoretically demonstrated.
  Our scheme relies on very elementary building blocks:
  nonrelativistic plane-wave electrons and laser fields, combined with quantum interference.
  It acts as a Stern-Gerlach device for free electrons, splitting the incident beam into spin components along the laser magnetic field direction.

  Free-electron laser facilities appear particularly promising for an experimental test of our predictions.
  Since the time scale for the spin-sensitive electron interaction is set by a rather slow bichromatic Rabi dynamics,
  low electron energies are advantageous to limit the required spatial extent of the laser fields.

  The following strengths of the setup are noteworthy:
  (i) Both outgoing partial beams are spin-polarized and separatly available for further use.
  (ii) It may equally well serve as spin filter and spin analyzer and also works with positrons.
  (iii) By proper synchronization, the scheme offers a novel method of creating and characterizing ultrashort (e.g., femtosecond) spin-polarized electron pulses.

  Ultrashort spin-polarized electron pulses would be relevant,
  e.g., to probe (de)magnetization phenomena \cite{Li2013, *Krieger2014} and chiral system dynamics
  \cite{[{See, e.g.\ the electron dynamics in chiral systems (ELCH) project, }] Elch}
  on femtosecond timescales,
  and to add an additional dimension to ultrafast electron diffraction \cite{McGregor2015, [{For an ultrafast electron diffraction experiment conducted at SLAC, see }]Weathersby2015} and microscopy \cite{Grillo2013, *Feist2015}.
  Thus, while the spin-polarizing beam splitter proposed here emanates from an almost 100-years old problem,
  it might develop into a useful tool, given the sustained progress in laser and electron-beam technology.

 \section{Acknowledgement}
  We thank S. Ahrens for stimulating discussions at the onset of this project.
  This study was supported by SFB TR18 of the German Research Foundation (DFG) under project No. B11.
 \input{dellweg.bbl}
\end{document}

%% file: dellweg.bbl
%